\documentclass[aps,prx,superscriptaddress,twocolumn,longbibliography,amsmath]{revtex4-2}

\usepackage{graphicx}
\usepackage{dcolumn}
\usepackage{bm}
\usepackage[hidelinks]{hyperref}

\usepackage[dvipsnames]{xcolor}
\newcommand{\AU}[1]{{\color{black}#1}}

\begin{document}


\title{Non-Majorana-origin of the half-integer conductance quantization elucidated by multi-terminal superconductor--quantum anomalous Hall insulator heterostructure}

\author{Anjana Uday}
\thanks{These authors contributed equally to this work}
\affiliation{Physics Institute II, University of Cologne, D-50937 K{\"o}ln, Germany}

\author{Gertjan Lippertz}
\thanks{These authors contributed equally to this work}
\affiliation{Physics Institute II, University of Cologne, D-50937 K{\"o}ln, Germany}

\author{Bibek Bhujel}
\affiliation{Physics Institute II, University of Cologne, D-50937 K{\"o}ln, Germany}

\author{Alexey A. Taskin} \email{taskin@ph2.uni-koeln.de}
\affiliation{Physics Institute II, University of Cologne, D-50937 K{\"o}ln, Germany}

\author{Yoichi Ando} \email{ando@ph2.uni-koeln.de}
\affiliation{Physics Institute II, University of Cologne, D-50937 K{\"o}ln, Germany}


\begin{abstract}
Chiral one-dimensional transport can be realized in thin films of a surface-insulating ferromagnetic topological insulator called quantum anomalous Hall insulator (QAHI). When superconducting (SC) pairing correlations are induced in the surface of such a material by putting an $s$-wave superconductor on the top, the resulting topological superconductivity gives rise to chiral Majorana edge-modes. A quantized two-terminal conductance of $\frac{1}{2}(e^2/h)$ was proposed as a smoking-gun evidence for the topological SC phase associated with a single chiral Majorana edge-mode. There have been experiments to address this proposal, but the conclusion remains unclear. Here, we formulate the edge transport in a multi-terminal superconductor--QAHI heterostructure using the Landauer-B\"uttiker formalism. Compared to the original proposal for the $\frac{1}{2}(e^2/h)$-quantization based on a simple two-terminal model, our formalism allows for deeper understanding of the origin of the quantization. The analysis of our experiments on multi-terminal devices unambiguously shows that the half-integer conductance quantization arises from the equilibration of the potentials of the incoming edge states at the SC electrode, and hence it is not of Majorana origin.
\end{abstract}


\maketitle

\newpage

\section{Introduction}
A thin film of a surface-insulating ferromagnetic topological insulator showing the quantum anomalous Hall effect (QAHE) \cite{Yu2010, Chang2013, Chang2015} is called quantum anomalous Hall insulator (QAHI). 
Inducing superconducting (SC) correlations in a QAHI through the SC proximity effect has been actively pursued in recent years, because it would lead to exotic topological superconductivity hosting one-dimensional (1D) chiral Majorana edge-modes \cite{Qi2010, Wang2015}. The creation of a $\pi$-phase domain boundary in these edge-modes is predicted to lead to mobile Majorana zero-modes, which could transfer quantum information between stationary topological qubits \cite{Beenakker2019a, Beenakker2019b, Adagideli2020}. Alternatively, if two counter-propagating chiral edge states of a QAHI is brought close together, either by etching a trench in the QAHI film or by etching the QAHI film into a nano-strip, then introduction of SC correlations between these two edges via the crossed Andreev reflection (CAR) using the SC proximity effect would lead to a quasi-1D topological superconductor with a pair of Majorana zero-modes at the ends \cite{Chen2018, Legendre2024, Clarke2014, Lee2017}. Hence, the proximitized QAHI constitutes an interesting platform for Majorana physics.

The QAHE showing the Hall resistance quantized to $h/e^2$ with vanishing longitudinal resistance is realized by doping ultra-thin films of the three-dimensional (3D) TI material (Bi$_x$Sb$_\text{1-x}$)$_2$Te$_3$ with Cr or V, and fine-tuning the composition $x$ to move the chemical potential into the magnetic gap opened at the Dirac point of the surface states as a result of the ferromagnetic order \cite{Yu2010, Chang2013, Chang2015}. No conclusive evidence was reported for the SC proximity effect in a QAHI~\cite{Kayyalha2020, Shen2020, He2017, Thorp2022, Atanov2024} until the recent observation of a negative resistance due to CAR across a narrow Nb finger electrode on top of a V-doped (Bi$_x$Sb$_\text{1-x}$)$_2$Te$_3$ thin film by our group \cite{Uday2024}. The negative nonlocal potential in the downstream edge stemming from the holes created as a result of the CAR process was observed only when the width of the Nb electrode was less than $\sim$500~nm. Since this experiment had a long finger-shaped SC electrode which promotes the CAR process beneath the finger, it is an interesting question if the negative resistance of different origin, such as Majorana interference \cite{Fu2009, Akhmerov2009} or Andreev edge-state transport \cite{Zhao2020, Hatefipour2022}, could be observed over a longer length scale in a differently-shaped electrode as a signature of SC proximity effect. 

\begin{figure}
\centering
\includegraphics[width=.48\textwidth, trim={0cm 0cm 0cm 0cm},clip]{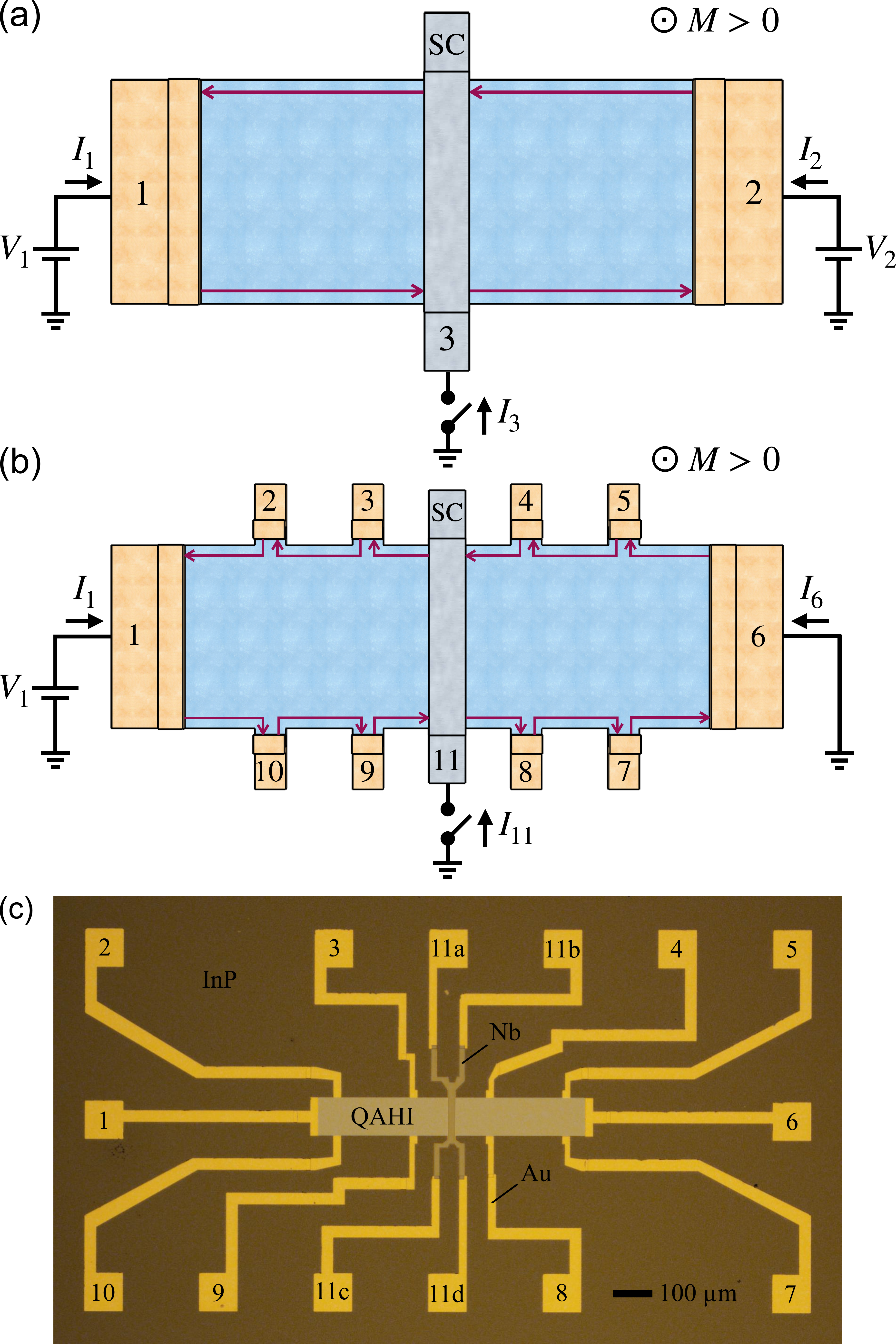}
\caption{(a) Simple two-terminal set-up on a QAHI thin film, in which contacts 1 and 2 are normal-metal contacts and the SC contact 3 can be grounded. (b) Schematic for a multi-terminal Hall-bar with the SC electrode 11 lying across the center of the device. Contacts 1 to 10 are normal-metal contacts. Current can flow into the QAHI film through contacts 1, 6, and 11. The other contacts are floating and serve as voltage probes. The chiral edge state of the QAHI is represented by the red arrows and shown for an upward, out-of-plane magnetization ($M > 0$). (c) False-color image of the measured multi-terminal Hall-bar device A with a SC Nb electrode lying across the center of the V-doped (Bi$_x$Sb$_\text{1-x}$)$_2$Te$_3$ thin film. The overlap of the superconductor with the QAHI thin film is $20 \times 100$~$\mu$m$^2$.}
\label{fig:setup}
\end{figure}

Another possible signature of the SC proximity effect in a QAHI is the quantized two-terminal conductance of $\frac{1}{2}(e^2/h)$ \cite{Wang2015}: If a topological SC phase with only one Majorana edge-mode ($\mathcal{N} = 1$) \cite{Qi2010} is realized as a result of the SC proximity effect underneath a SC strip lying across the full width of a QAHI [Fig.~\ref{fig:setup}(a)], one chiral Majorana edge-mode is transmitted longitudinally across the SC strip and the other transversely to the counter-propagating edge state on the other side of the sample, leading to the conductance of $\frac{1}{2}(e^2/h)$ \cite{Chung2011, Wang2015}. This reduction in two-terminal conductance by a factor of two, as compared to $e^2/h$ for a bare QAHI without a SC strip, was experimentally observed \cite{He2017, Thorp2022,  Huang2024, Kayyalha2020}, but its origin is still under intense debate as other trivial mechanisms were proposed \cite{Chen2017,Ji2018, Huang2018, Lian2018, Kayyalha2020, Huang2024}.


In the present work, we use the Landauer-B\"uttiker formalism to reevaluate the proposed half-integer-quantized two-terminal conductance as a signature of the SC proximity effect in a QAHI hall-bar with a $\mu$m-size SC electrode lying across the width of the device. With the formula derived for a multi-terminal device made on such a structure [Fig.~\ref{fig:setup}(b)], we analyze our experimental results on thin films of V-doped (Bi$_x$Sb$_\text{1-x}$)$_2$Te$_3$ proximitized by $\mu$m-size Nb superconducting electrodes. Rather than to only characterize the two-terminal conductance, we individually determine the potentials of all the chiral edge states in our multi-terminal devices. We show unambiguously that the half-integer-quantized two-terminal conductance arises from the edge state equilibration of the two chiral edge states arriving at the SC electrode (in agreement with Ref.~\cite{Kayyalha2020}). This is a trivial effect which also occurs in the absence of superconductivity. Lastly, no negative nonlocal edge potentials are observed in our devices, suggesting that signatures of the SC proximity effect in proximitized QAHI films can only be observed within the length scale of the SC coherence length, unlike the case of NbTiN-InAs heterostrutures in the quantum Hall insulator (QHI) regime which employs SC electrodes of similar (or even larger) sizes \cite{Hatefipour2022}.

\section{Half-integer conductance quantization}

It is instructive to revisit the original prediction of the half-integer quantization of the two-terminal conductivity for a QAHI Hall-bar with an $s$-wave superconductor strip lying across the full width of the device \cite{Chung2011, Wang2015}, as shown in Fig.~\ref{fig:setup}(a). To allow for Andreev scattering within the linear-response Landauer-B\"uttiker (LB) formalism \cite{Datta1995, Lambert1998}, the current-voltage relation can be written as
\begin{equation}
    I_i=\sum_{j=1}^3 a_{i j}\left(V_j-V_\text{SC}\right),
    \label{eq:LB_Ii}
\end{equation}
where $I_i$ is the single-particle current flowing into contact~$i$, $V_j$ is the potential at contact $j$, and $V_\text{SC} = V_3$ is the potential of the SC electrode (contact 3). The proportionality coefficients $a_\text{ij}$ in Eq.~\eqref{eq:LB_Ii} at zero temperature are given by 
\begin{equation}
    a_{i j}=\frac{e^2}{h}\left(N_i^{\mathrm{e}} \delta_{i j}-T_{i j}^{\mathrm{ee}}+T_{i j}^{\mathrm{eh}}\right),
    \label{eq:LB_aij}
\end{equation}
where $N_i^{\mathrm{e}}$ is the number of available channels for electron-like excitation in contact $i$, and $T_\text{ij}^\mathrm{ee}$ $(T_\text{ij}^\mathrm{eh})$ is the transmission coefficients of an electron from the $j$-th contact to arrive as an electron (hole) at the $i$-th contact. For the set-up shown in Fig.~\ref{fig:setup}(a), the relevant transmission coefficients are then: 
\begin{align*}
    T_{1,1}^\mathrm{ee} &= T_{2,2}^\mathrm{ee} \equiv T^\mathrm{ee}_\mathrm{T}, \\
    T_{1,1}^\mathrm{eh} &= T_{2,2}^\mathrm{eh} \equiv T^\mathrm{eh}_\mathrm{T}, \\
    T_{1,2}^\mathrm{ee} &= T_{2,1}^\mathrm{ee} \equiv T^\mathrm{ee}_\mathrm{L}, \\
    T_{1,2}^\mathrm{eh} &= T_{2,1}^\mathrm{eh} \equiv T^\mathrm{eh}_\mathrm{L}, \\
    T_{1,3}^\mathrm{ee} = T_{3,1}^\mathrm{ee} &= T_{2,3}^\mathrm{ee} = T_{3,2}^\mathrm{ee} \equiv T^\mathrm{D},
\end{align*}
with $T^\mathrm{ee}_\mathrm{T} + T^\mathrm{eh}_\mathrm{T} + T^\mathrm{ee}_\mathrm{L} + T^\mathrm{eh}_\mathrm{L} + T^\mathrm{D} = 1$, where the subscripts `L' and `T' refer to the transmission of a particle longitudinally underneath the SC strip and transversely across the width of the Hall-bar along the SC strip, respectively. $T^\mathrm{ee}_\mathrm{L}$ and $T^\mathrm{ee}_\mathrm{T}$ describe all the processes for an electron arriving at the SC electrode with an energy smaller than the SC gap to leave as an electron in one of the two edge states originating from the superconductor; these processes include: electron co-tunneling (CT), the transmission through chiral Andreev edge states (CAESs), and the transmission through chiral Majorana edge-modes (CMEMs). $T^\mathrm{eh}_\mathrm{L}$ and $T^\mathrm{eh}_\mathrm{T}$ describe all the processes that result in an electron arriving at the SC electrode with an energy smaller than the SC gap to leave as a hole in one of the two edge states originating from the superconductor; these processes include: CAR, CAESs and CMEMs. Moreover, $T^\mathrm{D}$ is included to describe single electrons entering the SC electrode with an energy smaller than the SC gap though subgap states, e.g.~via the interaction with vortices or due to the presence of a soft gap.

Using Eqs.~\eqref{eq:LB_Ii}-\eqref{eq:LB_aij}, it is then easy to show that:
\begin{equation}
    \frac{I_1-I_2}{V_1-V_2} = \frac{e^2}{h} (1 + k).
    \label{eq:G2T_e2_h}
\end{equation}
with
\begin{equation}
    k \equiv (T^\mathrm{ee}_\mathrm{L} - T^\mathrm{eh}_\mathrm{L}) - (T^\mathrm{ee}_\mathrm{T} - T^\mathrm{eh}_\mathrm{T}).
    \label{eq:k}
\end{equation}
Notice that within the expression for $k$ the $T^\mathrm{ee}_i$ and $T^\mathrm{eh}_i$ coefficients are competing with each other, as well as the longitudinal and transverse processes.

For a floating SC electrode ($I_1 = -I_2$), or when the voltage is applied symmetrically ($V_1 = -V_2$) with respect to a grounded SC electrode ($V_3 = 0$), the expression for the two-terminal (2T) conductance becomes:
\begin{equation}
    \sigma_\text{2T} = \frac{I_1}{V_1-V_2} = \frac{e^2}{2h}(1+k).
    \label{eq:G2T}
\end{equation}
The fact that Eq.~\eqref{eq:G2T} only depends on the parameter $k$ is the first disadvantage of using the two-terminal conductance to characterize the SC proximity effect; it is not possible to extract the individual transmission coefficients.

In the original proposals \cite{Chung2011, Wang2015}, Chung \textit{et al.}~predicted that for a proximitized QAHI in the $\mathcal{N} = 1$ topological SC state, the transmission coefficients obey the constraint $T^\mathrm{ee}_\mathrm{T} = T^\mathrm{eh}_\mathrm{T} = T^\mathrm{ee}_\mathrm{L} = T^\mathrm{eh}_\mathrm{L}$, as one chiral Majorana edge-mode is transmitted underneath the SC electrode and the other across the width of the Hall-bar along the SC strip. For the $\mathcal{N} = 2$ topological SC state with two CMEMs, they predicted $T^\mathrm{ee}_\mathrm{T} = T^\mathrm{eh}_\mathrm{T} = T^\mathrm{eh}_\mathrm{L} = 0$, as both chiral Majorana edge-modes are transmitted underneath the SC electrode. The expressions for the two-terminal conductance (Eq.~\eqref{eq:G2T}) then become:
\begin{align}
    \sigma_\text{2T} &= \frac{e^2}{2h} &&\mathrm{for} \quad \mathcal{N} = 1, \label{eq:G2T_N=1} \\
    &= \frac{e^2}{2h}(2 - T^\mathrm{D})  &&\mathrm{for} \quad \mathcal{N} = 2.\label{eq:G2T_N=2}
\end{align}
Chung \textit{et al.}~did not include $T^\mathrm{D}$ in their original model \cite{Chung2011, Wang2015}, resulting in perfect half-integer and integer quantization of $\sigma_\text{2T}$ for $\mathcal{N} = 1$ and $\mathcal{N} = 2$, respectively. However, if the single-particle current into the SC electrode is large ($T^\mathrm{D} \approx 1$), the $\mathcal{N} = 2$ topological SC state will also give $\sigma_\text{2T} = e^2/(2h)$.

It is important to point out that: (i) This half-integer $\sigma_\text{2T}$ signature is not unique to the above mentioned choice of transmission coefficients, and can show up for many combinations of $T^\mathrm{ee}_\mathrm{L}$, $T^\mathrm{eh}_\mathrm{L}$, $T^\mathrm{ee}_\mathrm{T}$, and $T^\mathrm{eh}_\mathrm{T}$. (ii) The situation where $T^\mathrm{ee}_\mathrm{L} = T^\mathrm{eh}_\mathrm{L}$ and $T^\mathrm{ee}_\mathrm{T} = T^\mathrm{eh}_\mathrm{T}$ corresponds to the QAH edge states leaving the SC electrode as an equal superposition of electron and hole, carrying a potential equal to the chemical potential of the SC electrode. As pointed out before \cite{Lee2017, Kayyalha2020}, this does not require chiral Majorana edge-modes and occurs naturally for chiral Andreev edge states traveling along a SC electrode over a long distance; for example in the case of a superconductor-QHI heterostructure over many skipping orbits. (iii) If the single-particle current into the SC electrode is large ($T^\mathrm{D} \approx 1$), then the two-terminal conductance is always $\sigma_\text{2T} \approx e^2/2h$. In this case the superconductor is indistinguishable from a normal metal contact. Hence, the half-integer quantization of the two-terminal conductivity is \textit{not} a smoking gun evidence for the $\mathcal{N} = 1$ topological SC state in a proximitized QAHI.

\begin{figure*}
\centering
\includegraphics[width=\textwidth, trim={0cm 0cm 0cm 0cm},clip]{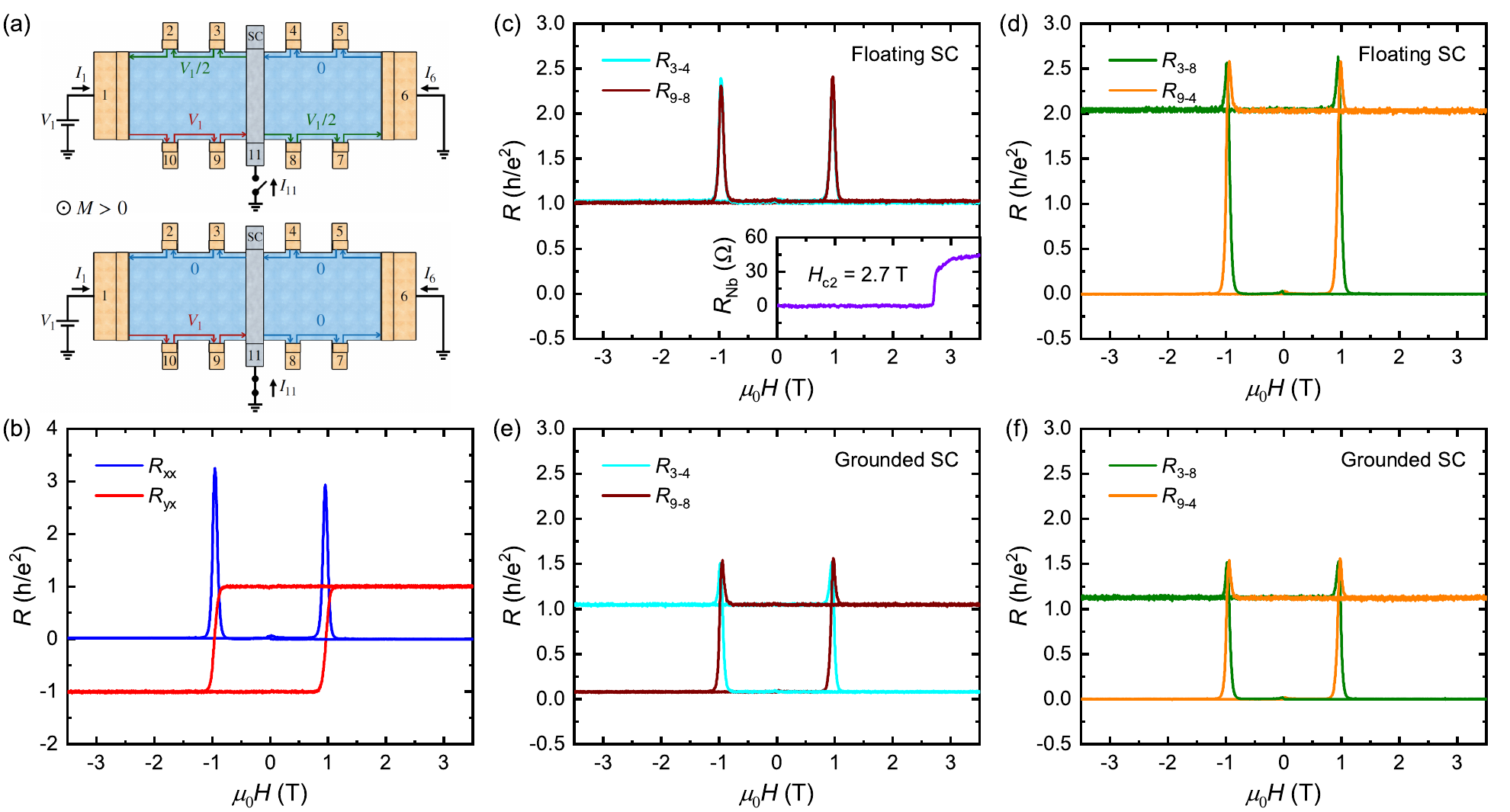}
\caption{(a) Schematic of the floating and grounded configurations, with the allowed edge potentials indicated near the arrows to show the edge states. (b) Longitudinal (transverse) resistance $R_{xx}$ ($R_{yx}$) at 65~mK for device A, showing a vanishing (quantized) value indicative of a clean QAHE realized in our device even after fabrication. (c-f) The magnetic-field dependencies of the resistances $R_{i\text{-}j} \equiv (V_i-V_j)/I_1$ at 65~mK for device A, measured between contacts i and j divided by the total current flowing into contact 1 ($I_1 = 1$~nA), for the floating (c-d) and grounded (e-f) configurations. From these resistance values, the edge potentials shown in (a) can be inferred. Inset of (c): Nb-electrode resistance vs external magnetic field, showing the upper critical field $H_\text{c2} \approx 2.7$~T.}
\label{fig:Nb_Overview}
\end{figure*}

\section{Experiments on Multi-terminal Hall-bar Devices with a SC Strip}

Hatefipour \textit{et al.}~observed a negative downstream resistance with respect to a 150-$\mu$m-wide grounded SC electrode in NbTiN-InAs heterostructures in the quantum Hall regime \cite{Hatefipour2022}. This is a remarkably long length scale as compared to the SC coherence length of NbTiN, which poses the question whether negative nonlocal resistances can also be observed in proximitized QAHI heterostructures containing $\mu$m-size SC electrodes. For this purpose, we added additional voltage terminals to our devices, see Figs.~\ref{fig:setup}(b-c). By measuring the four-terminal resistances of our multi-terminal Hall-bar devices with SC strip, we can directly determine the potential of the downstream edge. This configuration also helps to better understand the cause of the $e^2/(2h)$ quantization. 

We will limit our discussion to the case of an upward, out-of-plane magnetization ($M > 0$) of the QAHI thin films, which corresponds to a counterclockwise motion of the chiral 1D edge state. Note that under time-reversal symmetry, the transmission coefficients change as $T_{ij}(M>0) = T_{ji}(M<0)$, which means the expressions for the resistances across SC electrode defined below change as: $R_\text{3-4}(M>0) = R_\text{9-8}(M<0)$ and $R_\text{3-4}(M<0) = R_\text{9-8}(M>0)$.

\subsection{Fabrication Details}

The QAHI samples used in this study are uniformly V-doped (Bi$_x$Sb$_\text{1-x}$)$_2$Te$_3$ thin films with a thickness of $\sim$8~nm, grown on InP (111)A substrates by molecular beam epitaxy (MBE) in an ultra-high vacuum (UHV) environment. The details of the growth were already published in Refs.~\cite{Lippertz2022, Uday2024}. Atomic layer deposition (ALD) at 80$^\circ$C (Ultratec Savannah S200), is used to cover the freshly grown films \textit{ex-situ} with a 4-nm-thick Al$_2$O$_3$ capping layer to avoid degradation in air. The Hall-bar devices are patterned using standard optical lithography techniques. The Nb/Au SC contacts (45/5~nm for device A and 90/5~nm for device B) and the Ti/Au normal metal contacts (5/45~nm) are defined using electron-beam lithography. Aluminum etchant (Transene, Type-D) at 50$^\circ$C is used to selectively remove the Al$_2$O$_3$ capping layer before the sputter-deposition of the Ti/Au and Nb/Au layers in UHV. All QAHI films in this work display a quantized $R_{yx}$ and vanishing $R_{xx}$ without the need of electrostatic gating, see Fig.~\ref{fig:Nb_Overview}(b). 

This fabrication process was previously shown to result in good proximitization of V-doped (Bi$_x$Sb$_\text{1-x}$)$_2$Te$_3$ thin films with a low contact resistance \cite{Uday2024}. Hence, it is reasonable to assume, that by following the same recipe for the fabrication of the $\mu$m-size Nb electrodes in this work, the V-doped (Bi$_x$Sb$_\text{1-x}$)$_2$Te$_3$ underneath is also well-proximitized.


\subsection{Floating SC Electrode}

First, we treat the experimental set-up in which the SC contact 11 is floating, and the current flows from contact 1 to 6, see Fig.~\ref{fig:setup}(b). This is the same set-up as used in Refs.~\cite{He2017, Kayyalha2020, Thorp2022, Huang2024}, which investigated $\sigma_\text{2T}$ for a superconductor-QAHI heterostructures. Solving Eq.~\eqref{eq:LB_Ii} with \AU{$I_1 = -I_6$} together with Eqs.~\eqref{eq:I_zero}-\eqref{eq:I_nonzero} from the Appendix, yields
\begin{equation}
    V_{11} = \frac{V_1+V_6}{2} = \frac{V_2+V_7}{2}
    \label{eq:V11_float}
\end{equation}
independent of the choice of $T^\mathrm{D}$, $T^\mathrm{ee}_\mathrm{L}$, $T^\mathrm{eh}_\mathrm{L}$, $T^\mathrm{ee}_\mathrm{T}$, and $T^\mathrm{eh}_\mathrm{T}$. Equation~\eqref{eq:V11_float} shows that when the SC strip is floating, $V_{11}$ equilibrates the incoming edge state potentials ($V_1$ and $V_6$), but the outgoing chiral edge states need not be at the same potential ($V_2$ and $V_7$).

The resistances across the SC strip, $R_\text{3-4}$ and $R_\text{9-8}$, then become:
\begin{equation}
    \frac{V_3-V_4}{I_1} = \frac{V_9-V_8}{I_1} = \frac{h}{e^2}\frac{(1-k)}{(1+k)},
    \label{eq:R_float_3-4}
\end{equation}
with $k$ given by Eq.~\eqref{eq:k}. Note that $R_\text{3-4}$ and $R_\text{9-8}$ cannot become negative, since $-1 \leq k \leq  1$.

Figure \ref{fig:Nb_Overview}(c-d) shows the magnetic-field dependencies of the resistances measured across the Nb strip for this floating configuration. The potentials of the chiral edge states leaving the SC electrode are then easily determined: $V_3 \approx V_8 \approx V_1/2$, as illustrated in Fig.~\ref{fig:Nb_Overview}(a, top). This corresponds to $k \approx 0$. Notice that $R_\text{3-4}$ and $R_\text{9-8}$ remain unchanged when the magnetic field is swept (even for $H > H_\mathrm{c2}$), with the exception of the peaks at the coersive field where the sample's magnetization inverts. This means that if the sample undergoes a $\mathcal{N} = 1$ to $\mathcal{N} = 2$ topological SC phase transition, it cannot be seen in the transport data. Note that the value of $R_\text{3-4} \approx R_\text{9-8} \approx h/e^2$ corresponds to a two-terminal conductance $\sigma_\text{2T} \approx e^2/(2h)$ for the sample.

The observation of $k \approx 0$ can have two origins, i.e.~either the single-particle current into the superconductor is large ($T^\mathrm{D} \approx 1$), or the transmission coefficients are $T^\mathrm{ee}_\mathrm{T} \approx T^\mathrm{eh}_\mathrm{T}$ and $T^\mathrm{ee}_\mathrm{L} \approx T^\mathrm{eh}_\mathrm{L}$ as the CMEMs (or CAESs) at the superconductor-QAHI interface become an equal superposition of electron and hole for such large SC electrodes. In both cases, the superconductor essentially acts as a good metal contact, equilibrating the potentials of the incoming edge states. This is in agreement with the earlier report by Kayyalha \textit{et al.}~\cite{Kayyalha2020}. 
\AU{In passing, the concept of edge-state equilibration also plays an important role \cite{Protopopov2017, Nosiglia2018, Manna2024A, Manna2024B, Pandey2024} in understanding the peculiar conductance quantization across a quantum point contact made on the $\nu = 2/3$ fractional quantum Hall state \cite{Nakamura2023, Fauzi2023}.}

\subsection{Grounded SC Electrode}

Next we ground both the SC contact 11 and the metallic contact 6 (i.e. $V_{11} = V_6 = 0$). This set-up is similar to the nonlocal measurement of the downstream resistance performed with respect to a grounded SC electrode on top of (fractional) QHI and QAHI thin films \cite{Lee2017, Zhao2020, Gul2022, Hatefipour2022, Uday2024}. The resistances across the SC strip then become:
\begin{align}
    R_\text{3-4} &= \frac{V_3-V_4}{I_1} = \frac{h}{e^2}\frac{(T^\mathrm{ee}_\mathrm{T}-T^\mathrm{eh}_\mathrm{T})}{(1-T^\mathrm{ee}_\mathrm{T}+T^\mathrm{eh}_\mathrm{T})}, \label{eq:R_ground_3-4} \\
    R_\text{9-8} &= \frac{V_9-V_8}{I_1} = \frac{h}{e^2}\frac{(1 - T^\mathrm{ee}_\mathrm{L}+T^\mathrm{eh}_\mathrm{L})}{(1-T^\mathrm{ee}_\mathrm{T}+T^\mathrm{eh}_\mathrm{T})}, \label{eq:R_ground_9-8}
\end{align}
where $R_\text{3-4}$ is negative when $T^\mathrm{eh}_\mathrm{T} > T^\mathrm{ee}_\mathrm{T}$. For this grounded configuration, the resistances no longer depend on the parameter $k$. By combining various measurement configurations, one can now determine the amplitudes of $T^\mathrm{ee}_\mathrm{L}$, $T^\mathrm{eh}_\mathrm{L}$, $T^\mathrm{ee}_\mathrm{T}$, and $T^\mathrm{eh}_\mathrm{T}$ independently. Notice that $R_\text{3-4}$ is sensitive to the transverse transmission coefficients.

To probe the longitudinal transmission coefficients $T^\mathrm{ee}_\mathrm{L}$ and $T^\mathrm{eh}_\mathrm{L}$, it is useful to measure the resistance $R_\text{8-4}$ given by
\begin{equation}
    R_\text{8-4} = \frac{V_8-V_4}{I_1} = \frac{h}{e^2}\frac{(T^\mathrm{ee}_\mathrm{L}-T^\mathrm{eh}_\mathrm{L})}{(1-T^\mathrm{ee}_\mathrm{T}+T^\mathrm{eh}_\mathrm{T})},
    \label{eq:R_ground_8-4}
\end{equation}
which becomes negative when $T^\mathrm{eh}_\mathrm{L} > T^\mathrm{ee}_\mathrm{L}$; note that this $R_\text{8-4}$ is different from the usual transverse resistance $R_{yx} = -(V_8-V_4)/I_6 = +h/e^2$, see Eq.~\eqref{eq:Ryx} in the Appendix. Hence, the grounded SC configuration allows for a straightforward differentiation between ($T^\mathrm{ee}_\mathrm{L}$, $T^\mathrm{eh}_\mathrm{L}$) and ($T^\mathrm{ee}_\mathrm{T}$, $T^\mathrm{eh}_\mathrm{T}$), unlike the floating superconductor configuration where all the resistances only depend on $k$ (Eq.~\eqref{eq:k}). 

Figures \ref{fig:Nb_Overview}(e-f) show the magnetic-field dependencies of the resistances measured across the Nb strip for this grounded configuration. The potentials of the chiral edge states leaving the SC electrode are found to be $V_3 \approx V_8 \approx V_6 = 0$, as illustrated in Fig.~\ref{fig:Nb_Overview}(a, bottom). This corresponds again to the Nb strip equilibrating the potentials of the incoming edge states. However, notice that $R_\text{3-4}(M > 0)$ and $R_\text{9-8}(M < 0)$ in Fig.~\ref{fig:Nb_Overview}(e) are now nonzero and positive, respectively. Similarly, $R_\text{3-8}(M > 0)$ and $R_\text{9-4}(M < 0)$ in Fig.~\ref{fig:Nb_Overview}(f) deviate strongly from $h/e^2$. Rather than to attribute this to $T^\mathrm{ee}_\mathrm{T} > T^\mathrm{eh}_\mathrm{T}$ and $T^\mathrm{ee}_\mathrm{L} > T^\mathrm{eh}_\mathrm{L}$, we interpret this to be an artifact of nonideal contacts. 
In an actual experiment, both contacts 11 and 6 will always have a finite contact resistance to the QAHI film. This will cause the current to be divided between the two grounds [see Eq. ~\eqref{eq:I_ground_nonideal} in the appendix]. In device A, we estimate $\sim$8\% of the total current flew through contact 6.

Regardless of the above mentioned artifact, the resistances in Figs.~\ref{fig:Nb_Overview}(e-f) do not change as the magnetic field is increased above the upper critical field of Nb $H_\mathrm{c2} = 2.7$~T [see inset of Figs.~\ref{fig:Nb_Overview}(c)]. This leads us to conclude that $T^\mathrm{ee}_\mathrm{T} \approx T^\mathrm{eh}_\mathrm{T}$ and $T^\mathrm{ee}_\mathrm{L} \approx T^\mathrm{eh}_\mathrm{L}$.

\begin{figure}
\centering
\includegraphics[width=.40\textwidth, trim={0cm 0cm 0cm 0cm},clip]{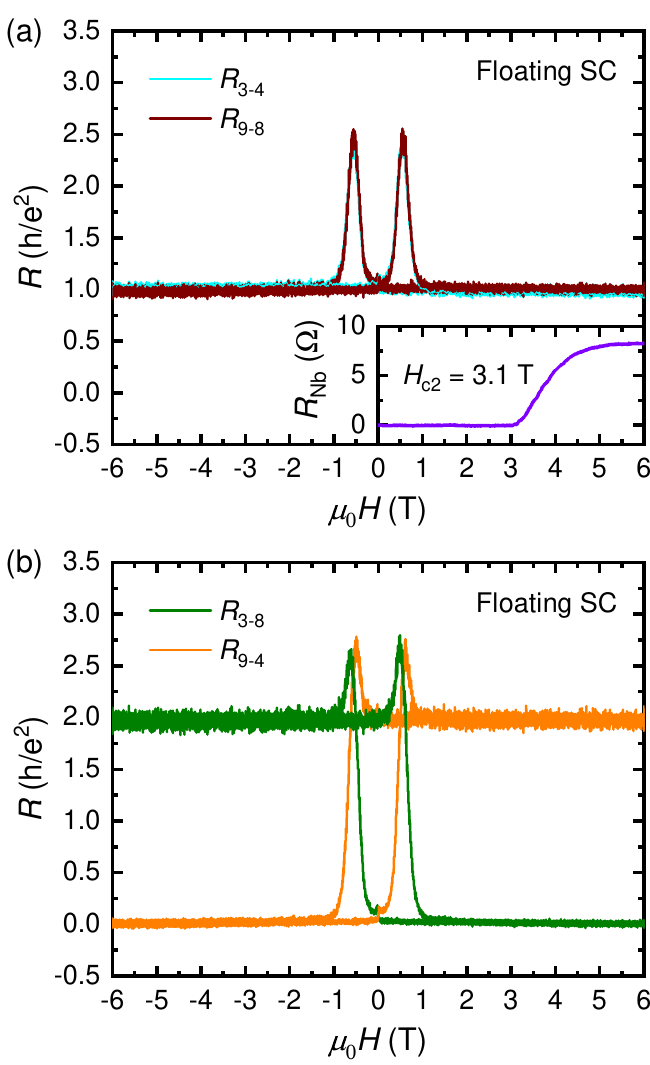}
\caption{(a-b) The magnetic-field dependencies of the resistances $R_{i\text{-}j} \equiv (V_i-V_j)/I_1$ at 65~mK for device B in the floating configuration. In this device, the QAHI film is interrupted by a 10-$\mu$m-wide gap underneath the SC electrode, creating two hall-bar regions connected in series via the Nb strip. Inset of (a): Nb-electrode resistance vs external magnetic field, showing the upper critical field $H_\text{c2} \approx 3.1$~T.}
\label{fig:Nb_Trench}
\end{figure}

\subsection{Floating SC Electrode with Trench}

It is important to investigate whether or not the observation of $T^\mathrm{ee}_\mathrm{T} \approx T^\mathrm{eh}_\mathrm{T}$ and $T^\mathrm{ee}_\mathrm{L} \approx T^\mathrm{eh}_\mathrm{L}$ in our experiment could correspond to the $\mathcal{N} = 1$ topological SC state, in which one chiral Majorana edge-mode is transmitted underneath the SC electrode and the other across the width of the Hall-bar along the SC strip \cite{Chung2011, Wang2015}. We fabricated a second Hall-bar device where the QAHI thin film is interrupted by a 10-$\mu$m-wide gap underneath the SC electrode ($30 \times 100$~$\mu$m$^2$). This ensures that $T^\mathrm{ee}_\mathrm{L} = T^\mathrm{eh}_\mathrm{L} = 0$ and no chiral Majorana mode can be transmitted. All the current flowing from contact 1 to 6 has to go through the SC electrode now. Nevertheless, we still find the same edge potentials for the floating configuration: $V_3 = (1-k)V_1/2 \approx V_1/2$ and $V_8 = (1+k)V_1/2 \approx V_1/2$, as can be deduced from Fig.~\ref{fig:Nb_Trench}. Hence, $k = T^\mathrm{eh}_\mathrm{T} - T^\mathrm{ee}_\mathrm{T} \approx 0$ for this device. This means that the edge potentials remain unchanged regardless of how the contact is made to the SC electrode.

\subsection{Absence of Negative Nonlocal Resistance}

Lastly, we would like to comment on possible negative downstream potentials measured with respect to the floating SC electrode, as the observation of negative potentials is complicated in the grounded configuration due to the presence of two grounded contacts. A negative potential can, in principle, be observed for $V_2$ or $V_7$ as an offset from the equilibrated potential $V_{11} = (V_1+V_6)/2$ of the floating superconductor, see Eq.~\eqref{eq:V11_float}. The resistances measured with respect to the floating SC contact 11 then become:
\begin{align}
    R_\text{3-11} &= -R_\text{8-11} = -\frac{h}{e^2}\frac{k}{(1+k)} \label{eq:R_float_3-11}, \\
    R_\text{4-11} &= -R_\text{9-11} = -\frac{h}{e^2}\frac{1}{(1+k)} \label{eq:R_float_4-11},
\end{align}
where $R_\text{3-11}$ or $R_\text{8-11}$ can become negative depending on $k$ (Eq.~\eqref{eq:k}). Notice that $R_\text{8-11}$ (and $R_\text{4-11}$) are defined in the direction opposite to the current flow. This means that at elevated temperature or high bias current the non-zero longitudinal resistance can cause $R_\text{8-11}$ to become negative as well. The measurement set-up used for Eqs.~\eqref{eq:R_float_3-11}-\eqref{eq:R_float_4-11} has an advantage over the set-up used in Refs.~\cite{Lee2017, Zhao2020, Gul2022, Hatefipour2022, Uday2024}, where the SC electrode was grounded. Namely, when performing the experiment on a real device there will be no extrinsic contact resistance contribution to the resistances in Eqs.~\eqref{eq:R_float_3-11}-\eqref{eq:R_float_4-11}, as they are measured in a four-terminal configuration.

\begin{figure}
\centering
\includegraphics[width=.4\textwidth, trim={0cm 0cm 0cm 0cm},clip]{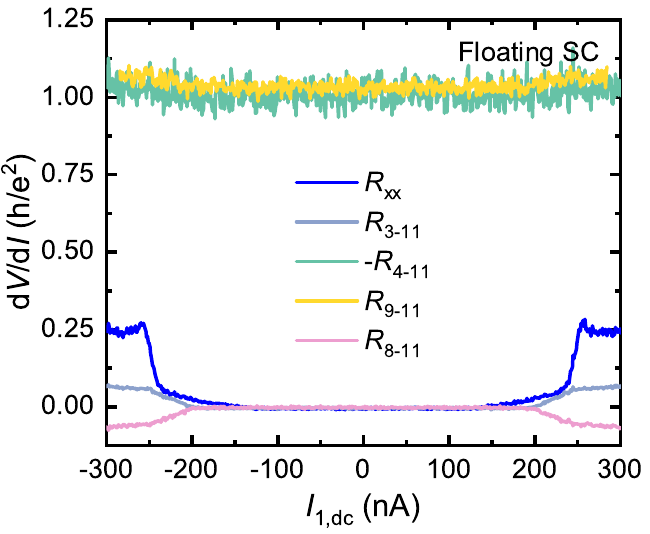}
\caption{Differential resistances at 17~mK measured as a function of the dc bias current flowing into contact 1 with a small ac excitation of 1~nA, for device A in the floating configuration. The measurement set-up is shown in Fig.~\ref{fig:Nb_Overview}(a). The resistances $R_\text{3-11}$ and $R_\text{8-11}$ are zero at low dc bias, meaning that there are no signatures of CAR or CT processes. At high bias, the resistances become dominated by the breakdown of the QAHE at $I_\text{BD} \approx 235$~nA \cite{Lippertz2022}, with already a slight onset of dissipation at $\sim$140~nA. The sample was magnetized at $+1.5$~T before performing the d$V$/d$I$ measurements at 0~T.}
\label{fig:Nb_dVdI}
\end{figure}

Figure \ref{fig:Nb_dVdI} shows the differential resistances measured with respect to the floating SC electrode 11, together with the longitudinal resistance $R_{xx}$, as a function of the dc bias current. The sample remains in the zero-resistance state up to $\sim$140~nA, after which there is an onset of dissipation in $R_{xx}$ in the pre-breakdown regime \cite{Fox2018}. At $I_\text{BD} \approx 235$~nA the sharp increase in $R_{xx}$ signifies the breakdown of the QAHE \cite{Kawamura2017, Fox2018, Lippertz2022, Qiu2022, Zhou2023, Fijalkowski2024, Roeper2024}, likely due to the electric-field-driven percolation of charge puddles across the width of the sample \cite{Lippertz2022}. As a result, the negative value of $R_\text{8-11} (M>0)$ stems from the nonzero longitudinal resistance at high dc bias current. At low bias, before the onset of dissipation, $R_\text{8-11} (M>0)$ is zero in Fig.~\ref{fig:Nb_dVdI}. Hence, no negative downstream resistances are observed when a V-doped (Bi$_\text{x}$Sb$_\text{1-x}$)$_2$Te$_3$ QAHI is proximitized by a $\mu$m-size SC electrode on top of the film.

This means that, unlike the case of a superconductor-QHI system based on InAs \cite{Hatefipour2022}, the signatures of SC proximity effect in QAHI require a SC electrode with dimensions of the order of the (induced) coherence length. When it is satisfied, $T^\mathrm{eh}_\mathrm{L}$ and $T^\mathrm{eh}_\mathrm{T}$ correspond to the probabilities of crossed Andreev reflection at the SC electrode in the longitudinal and transverse directions, respectively. The corresponding electron co-tunneling processes are then described by $T^\mathrm{ee}_\mathrm{L}$ and $T^\mathrm{ee}_\mathrm{T}$, respectively. In this regard, the recent sub-$\mu$m-size Hall-bars of QAHI reported in Ref.~\cite{Qiu2022, Zhou2023} may be a good platform to observe negative nonlocal resistances, if SC electrodes are fabricated with dimensions comparable to the (induced) SC coherence length \cite{Uday2024}.

\section{Conclusion}
We have shown that the proposed detection of the chiral Majorana edge-mode in a superconductor-QAHI heterostructure through the observation of a two-terminal conductance of $e^2/(2h)$ is ill-conceived. This conclusion was derived by formulating a Landauer-B\"uttiker model for the relevant multi-terminal set-up and performing experiments to elucidate the transmission coefficients, using  multi-terminal devices made of V-doped (Bi$_\text{x}$Sb$_\text{1-x}$)$_2$Te$_3$ thin films proximitized by Nb superconductor. Any experimental results on the two-terminal configuration can be explained by the SC electrode equilibrating all the chiral edge state potentials, possibly due to the presence of in-gap states at the superconductor-QAHI interface. While the sub-$\mu$m-size SC electrodes in our previous study showed clear evidence of the SC proximity effect in the form of negative downstream edge potentials \cite{Uday2024}, the chiral 1D edge states leaving from the $\mu$m-size SC electrodes in the present study always had a potential equal to that of the SC electrode. This shows that SC contacts on the order of the (induced) SC coherence length are required for the study of the SC proximity effect.

\section{Data availability}
\AU{The data used in the generation of the main and supplementary figures are available on Zenodo with the identifier \href{https://zenodo.org/records/14176676}{10.5281/zenodo.14176676} \cite{Zenodo}.}

\section{Acknowledgements}
This work has received funding from the Deutsche Forschungsgemeinschaft (DFG, German Research Foundation) under CRC 1238-277146847 (subprojects A04 and B01) and also from the DFG under Germany’s Excellence Strategy -- Cluster of Excellence Matter and Light for Quantum Computing (ML4Q) EXC 2004/1-390534769.

\section{Appendix}

\subsection{Transmission \& Proportionality Coefficients}

For an upward, out-of-plane magnetization ($M > 0$), the relevant transmission coefficients for the set-up shown in Fig.~\ref{fig:setup}(b) are:
\begin{gather*}
    T^\mathrm{ee}_{3,4} = T^\mathrm{ee}_{8,9} = T^\mathrm{ee}_\mathrm{L} \quad \mathrm{and} \quad T^\mathrm{eh}_{3,4} = T^\mathrm{eh}_{8,9} = T^\mathrm{eh}_\mathrm{L}, \\
    T^\mathrm{ee}_{3,9} = T^\mathrm{ee}_{8,4} = T^\mathrm{ee}_\mathrm{T} \quad \mathrm{and} \quad T^\mathrm{eh}_{3,9} = T^\mathrm{eh}_{8,4} = T^\mathrm{eh}_\mathrm{T}, \\
    T^\mathrm{ee}_{3,11} = T^\mathrm{ee}_{8,11} = T^\mathrm{ee}_{11,4} = T^\mathrm{ee}_{11,9} = T^\mathrm{D}.
\end{gather*}
Using Eq.~\eqref{eq:LB_aij}, the nonzero proportionality coefficients for $M > 0$ then become:
\begin{gather*}
    \begin{split}
    a_{1,1} &= -a_{1,2} = a_{2,2} = -a_{2,3} = a_{3,3} = a_{4,4} = -a_{4,5} \\
    &= a_{5,5} = -a_{5,6} =  a_{6,6} = -a_{6,7} = a_{7,7} = -a_{7,8} \\
    &= a_{8,8} = a_{9,9} = -a_{9,10} = a_{10,10} = -a_{10,1} = \frac{e^2}{h}, 
    \end{split} \\
    \begin{split}
    a_{3,4} = a_{8,9} &= \frac{e^2}{h} (T^\mathrm{eh}_\mathrm{L} - T^\mathrm{ee}_\mathrm{L}), \\
    a_{3,9} = a_{8,4} &= \frac{e^2}{h} (T^\mathrm{eh}_\mathrm{T} - T^\mathrm{ee}_\mathrm{T}), \\
    a_{11,11} &= \frac{2e^2}{h} T^\mathrm{D},
    \end{split} \\
    a_{3,11} = a_{8,11} = a_{11,4} = a_{11,9} = -\frac{e^2}{h} T^\mathrm{D}.
\end{gather*}
Note that contact 11 acts as both a metallic drain (through $T^\mathrm{D}$) and as a SC `scattering' object allowing electrons coming from contacts 4 and 9 to be transported via the CAR or CT processes to contacts 3 and 8 (through $T^\mathrm{eh}_\mathrm{L}$, $T^\mathrm{ee}_\mathrm{L}$, $T^\mathrm{eh}_\mathrm{T}$, and $T^\mathrm{ee}_\mathrm{T}$).

\subsection{Longitudinal \& Transverse Resistance}

Next, Eq.~\eqref{eq:LB_Ii} is solved (with the summation running over all eleven contacts) with $V_\mathrm{SC} = V_{11}$ and 
\begin{gather}
    I_2 = I_3 = I_4 = I_5 = I_7 = I_8 = I_9 = I_{10} = 0, \label{eq:I_zero} \\
    I_1 + I_6 + I_{11} + I^\mathrm{SC}_{11} = 0, \label{eq:I_nonzero}
\end{gather}
where $I_{11}$ is the single-particle current and $I^\mathrm{SC}_{11}$ is the supercurrent flowing into the device from contact 11. This yields the expressions for the longitudinal and transverse resistance:
\begin{gather}
    R_{xx} = \frac{V_2 - V_3}{I_1} = \frac{V_4 - V_5}{-I_6} = \frac{V_8 - V_9}{-I_6} = \frac{V_{10} - V_9}{I_1} = 0, \label{eq:Rxx} \\
    R_{yx} = \frac{V_{10} - V_2}{I_1} = \frac{V_9 - V_3}{I_1} = \frac{V_8 - V_4}{-I_6} = \frac{V_7 - V_5}{-I_6} = \frac{h}{e^2}, \label{eq:Ryx}
\end{gather}
as expected, regardless of whether the SC contact 11 is grounded or floating. The resistances measured across the SC strip ($R_\text{3-4}$ and $R_\text{9-8}$), on the other hand, do depend on whether (super)current is allowed to flow into the device through contact 11. These resistances are discussed for two different exprimental set-ups in the main text.

\subsection{Non-ideal Grounded Contacts}

For the measurements shown in Figs.~\ref{fig:Nb_Overview}(e-f), the SC contact 11 and the normal metal contact 6 are both grounded (i.e. $V_{11} = V_6 = 0$). In the actual experiment, both contact 11 and 6 have a finite contact resistance $R_{\mathrm{c},i}$ to the QAHI thin film, which means that the potentials $V_{11} = -I_{11} R_\mathrm{c,11}$ and $V_6 = -I_6 R_\mathrm{c,6}$ at the sample side of the contact will not be equal and slightly higher than zero. This will cause a small current to flow from contacts 11 to contact 6, which was not taken into account in Eqs.~\eqref{eq:R_ground_3-4}-\eqref{eq:R_ground_8-4}.

In the absence of CAR/CT and normal processes at the superconductor-QAHI interface, i.e.~$T^\mathrm{ee}_\mathrm{L} = T^\mathrm{eh}_\mathrm{L} = T^\mathrm{ee}_\mathrm{T} = T^\mathrm{eh}_\mathrm{T} = 0$ and $T^\mathrm{D} = 1$, we find:
\begin{equation}
    I_6 = \frac{R_\mathrm{c,11}}{R_\mathrm{c,6} + h/e^2} I_{11},
    \label{eq:I_ground_nonideal}
\end{equation}
which is a simple current divider. 

The resistances for the grounded configuration then become:
\begin{gather}
    R_\text{3-4} = R_\text{8-4} = \frac{R_\mathrm{c,11}}{1+(R_\mathrm{c,6} + R_\mathrm{c,11})e^2/h} \approx R_\mathrm{c,11}, \label{eq:R_ground_nonideal_3-4} \\
    R_\text{9-8} = \frac{h}{e^2}. \label{eq:R_ground_nonideal_9-8}
\end{gather}
Notice that $R_\text{3-4}$ and $R_\text{8-4}$ are positive and will possibly mask the negative resistances stemming from Andreev processes at the superconductor-QAHI interface.

\begin{figure}
\centering
\includegraphics[width=.4\textwidth, trim={0cm 0cm 0cm 0cm},clip]{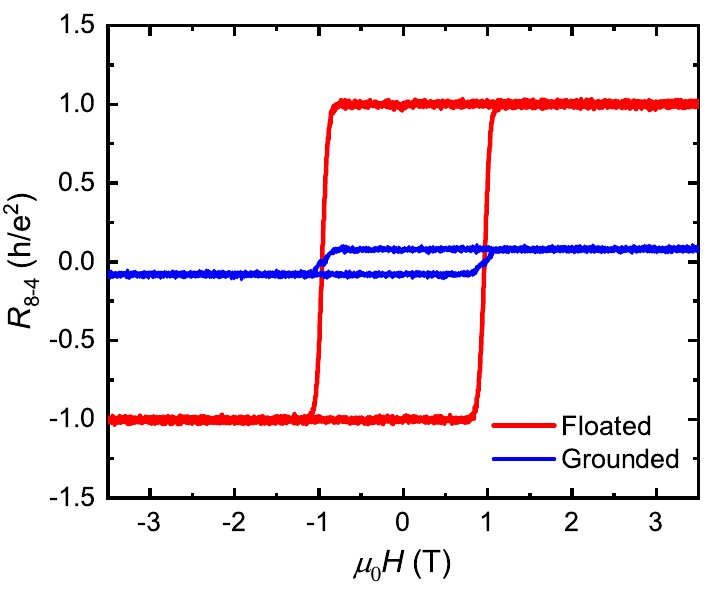}
\caption{The apparent transverse resistance $R_\text{8-4} = (V_8-V_4)/I_1$ with $I_1 = 1$~nA, for the floating and grounded configuration. The nonzero $R_\text{8-4}$ in the grounded configuration corresponds to $I_6 \approx -80$~pA which stems from the finite contact resistances of contacts 6 and 11 [see Eq.~\eqref{eq:R_ground_nonideal_3-4}].}
\label{fig:nonideal}
\end{figure}

Figure~\ref{fig:nonideal} shows $R_\text{8-4} = (V_8-V_4)/I_1$ for the floating and grounded configuration. When the SC electrode is floating then $I_1 = -I_6$ and $R_\text{8-4} = h/e^2$ to-within the accuracy of our measurement. On the other hand, when the SC electrode is grounded, $R_\text{8-4}$ in Fig.~\ref{fig:nonideal} is nonzero. This corresponds to a current $I_6 \approx -80$~pA flowing on the right side of the Nb strip. Rather than claiming that $T^\mathrm{ee}_\mathrm{L} > T^\mathrm{eh}_\mathrm{L}$ [Eq.~\eqref{eq:R_ground_8-4}], 
we attribute the nonzero $R_\text{8-4}$ to the presence of a finite contact resistance at the SC electrode 11, $R_\mathrm{c,11}$ [see Eq.~\eqref{eq:R_ground_nonideal_3-4}]. 
This is backed up by the fact that $R_\text{8-4}$ remains unchanged as the magnetic field is increased above the upper critical field of Nb, $H_\mathrm{c2} = 2.7$~T [see inset of Figs.~\ref{fig:Nb_Overview}(c)].

Note that the effect of non-ideal contacts for the grounded configuration in our experiment is large, as contacts 6 and 11 were grounded outside the dilution refrigerator (rather than on-chip). This means that in this case the `contact' resistance $R_\mathrm{c,i}$ also includes the line and filter resistances. Equation~\eqref{eq:R_ground_nonideal_3-4} gives a lower bound for $R_\mathrm{c,11} > R_\text{8-4} \approx 2$~k$\Omega$.

\subsection{Does a thin insulating barrier between the superconductor and QAHI help?}

In a recent publication \cite{Huang2024}, Huang \textit{et al.}~claimed that an AlO$_x$ oxide barrier suppresses the single-particle current into the SC electrode ($T^\mathrm{D} \approx 0$), whereas the SC proximity effect survives up to slightly higher barrier thicknesses. The observations of kinks at $\sigma_\text{2T} \approx 0.57\text{-}0.59\, \frac{e^2}{h}$ in the magnetic-field behaviour of the two-terminal conductance measured across a Nb strip were interpreted as signatures of the $\mathcal{N} = 1$ topological SC state [Eq.~\eqref{eq:G2T_N=1}]. In the supposed $\mathcal{N} = 2$ topological SC state, on the other hand, the samples showed $\sigma_\text{2T} \approx 0.74\, \frac{e^2}{h}$ [Eq.~\eqref{eq:G2T_N=2}].

The observations of Huang \textit{et al.} in Ref.~\citenum{Huang2024} are similar to those reported by He \textit{et al.} in the now-retracted work \cite{He2017, Thorp2022}, although the $\sigma_\text{2T}$ values are only within 10-30\% of the expected quantization. This was attributed to additional conduction channels and a remaining electric short through the Nb. Here, we examine whether these values can be reconciled with a high resistive short across the width of the Hall-bar. We assume $T^\mathrm{D} = 0$ to follow the claim of Ref.~\citenum{Huang2024}, and instead allow a fraction $T^\mathrm{S}$ of the current to flow between the opposing edges of the Hall-bar through the Nb to consider the electric short assumed in Ref.~\citenum{Huang2024}. The transmission coefficients for an electron to transmit as an electron between contacts 3, 4, 8, and 9 are then modified to:
\begin{equation*}
    T^\mathrm{ee}_{3,9} = T^\mathrm{ee}_{8,4} = T^\mathrm{ee}_\mathrm{T} + T^\mathrm{S},
\end{equation*}
and the new proportionality coefficients become:
\begin{gather}
    a_{3,9} = a_{8,4} = \frac{e^2}{h} (T^\mathrm{eh}_\mathrm{T} - T^\mathrm{ee}_\mathrm{T}-T^\mathrm{S}), \\
    a_{11,11} = a_{3,11} = a_{8,11} = a_{11,4} = a_{11,9} = 0.
\end{gather}

Using the condition $T^\mathrm{ee}_\mathrm{T} = T^\mathrm{eh}_\mathrm{T} = T^\mathrm{ee}_\mathrm{L} = T^\mathrm{eh}_\mathrm{L}$ for the $\mathcal{N} = 1$ topological SC state, and $T^\mathrm{ee}_\mathrm{T} = T^\mathrm{eh}_\mathrm{T} = T^\mathrm{eh}_\mathrm{L} = 0$ for the ${N} = 2$ topological SC state, the expressions for the two-terminal conductance then become:
\begin{align}
    \sigma_\text{2T} &= (1-T^\mathrm{S})\frac{e^2}{2h} &&\mathrm{for} \quad \mathcal{N} = 1, \label{eq:G2T_N=1_T^S} \\
        &= (1-T^\mathrm{S})\frac{e^2}{h}  &&\mathrm{for} \quad \mathcal{N} = 2. \label{eq:G2T_N=2_T^S}
\end{align}
Hence, if the $\mathcal{N} = 2$ topological SC state (which is indistinguishable from the QAHI state) yields $\sim 0.74\, \frac{e^2}{h}$ \cite{Huang2024}, we should search for a feature at $\sim 0.37\, \frac{e^2}{h}$ for the $\mathcal{N} = 1$ topological SC state, as one would also intuitively expect. As a result, the kinks observed by Huang \textit{et al.}~at $0.57\text{-}0.59\, \frac{e^2}{h}$ are most likely not related to the $\mathcal{N} = 1$ topological SC state, and furthermore, their study in Ref.~\cite{Huang2024} gives no evidence for induced superconductivity in a QAHI. 
Moreover, the magnetic-field dependencies of the two-terminal conductance reported in Ref.~\citenum{Huang2024} show several kinks at different values of $\sigma_\text{2T}$, which suggests that the kinks at $0.57\text{-}0.59\, \frac{e^2}{h}$ are not special points. Such features are possibly caused by temperature effects \cite{Bestwick2015} or inhomogeneous switching of the magnetization \cite{Lachman2015, Lachman2017}, affecting the longitudinal conductance of the QAHI film. 


%

\end{document}